\documentstyle[12pt,cite,epsf]{article}                                        
\catcode`\@=11                                                       
\begin{document}                                                     

\renewcommand{\theequation}{\thesection.\arabic{equation}}        
\newcommand{\mysection}[1]{\setcounter{equation}{0}\section{#1}}

\def\1{{\bf 1}}
\def\Z{{\bf Z}}
\def\ee{\end{equation}}
\def\be{\begin{equation}}
\def\l{\label}
\def\dxi{\partial_\xi}
\def\D{{\cal D}}
\def\sin{{\rm sin}}
\def\cos{{\rm cos}}
\def\f{{\bf \Phi}}
\def\v{\varphi}
\def\O{\bf {\cal O}}
\def\C{\bf C}
\def\CP{\bf CP}
\def\e{\rm e}
\def\0{\nonumber}
\def\eea{\end{eqnarray}}
\def\bea{\begin{eqnarray}}
\def\Tr{\rm Tr}
\def\IR{\bf R}
\def\ZZ{\bf Z}

%
%
\input epsf
\newcounter{fignum}
\newcommand{\figuurnum}{\arabic{fignum}}
\newcommand{\figuur}[2]{
\addtocounter{fignum}{1}
\addcontentsline{lof}{figure}{\protect
\numberline{\arabic{section}.\arabic{fignum}}{#2}}
\hspace{-3mm}{\it fig.}\ \figuurnum.
\begin{figure}[t]\begin{center}
\leavevmode\hbox{\epsffile{#1.eps}}\\[3mm]
\parbox{10cm}{\small \bf Fig.\ \figuurnum : \it #2}
\end{center} \end{figure}\hspace{-1.5mm}}
%
%
\newcommand{\figuurplus}[3]{
\addtocounter{fignum}{1}
\addcontentsline{lof}{figure}{\protect
\numberline{\arabic{section}.\arabic{fignum}}{#3}}
\hspace{-3mm}{\it Fig.}\ \figuurnum.
\begin{figure}[t]\begin{center}
\leavevmode\hbox{\epsfxsize=#2 \epsffile{#1.eps}}\\[5mm]
\parbox{10cm}{\small \bf Fig.\ \figuurnum: \it #3}
\end{center} \end{figure}\hspace{-1.5mm}}
\newcommand{\fig}{{\it fig.}\ }


\begin{titlepage}

\hfill{ULB-TH/02-15}

\vspace{2 cm}

\centerline{{\huge{Matrix Strings in pp-wave backgrounds}}}
\centerline{{\huge{from deformed Super Yang-Mills Theory}}}

\vspace{2 cm}

\centerline{Giulio Bonelli\footnote{e-mail address: gbonelli@ulb.ac.be}}
\centerline{Physique Theorique et Mathematique}
\centerline{Universite' Libre de Bruxelles}
\centerline{Campus Plaine C.P. 231; B 1050 Bruxelles, Belgium}
\centerline{and}
\centerline{Theoretische Natuurkunde}
\centerline{Vrije Universiteit Brussel}
\centerline{Pleinlaan, 2; B 1050 Brussel, Belgium}

\vspace{4 cm}

{\it Abstract}:
We formulate matrix models for strings in ten dimensional pp-wave backgrounds
and for particles in eleven dimensional ones.
This is done by first
characterizing the deformations of ten dimensional {\cal N}=1 SYM
which are induced by a constant bispinorial coupling $\int \bar\Psi H\Psi$
plus a minimal purely bosonic completition and then by the appropriate
dimensional reduction.
We find a whole class of new models for the matrix strings and 
a generalization of the supernumerary supersymmetric models as far as
the matrix theory for particles is concerned.
A companion deformation of the IKKT matrix model is also discussed.
\end{titlepage}

\section{Introduction}

Despite a lot of efforts, string theory in curved backgrounds remains 
a hard problem to be understood. This is because, generically, the 
perturbative methods apply only if the gravitational background
is almost flat (in string units) and a true nonperturbative
formulation of the theory is still to be given. 
The only partly positive results in this direction being some cases of homogeneous
spaces whose study is available because of the power of the
WZNW model (see for example \cite{NW}).

A simple case of tractable model is the string in the pp-wave \cite{mat}.
The pp-wave solution of IIB ten dimensional supergravity is given by
the metric
\be
ds^2=-4dx_+dx_--f^2{\bf x}^2d{x^+}^2+d{\bf x}^2
\l{IIBmet}\ee
where ${\bf x}$ is the eight dimensional coordinate vector,
and the self-dual constant RR 5-form
$$
F_{+1234}=F_{+5678}=f\times C
$$
where $C$ is a non zero numerical constant.
The light-cone GS string obtained in \cite{mat} is given by the quadratic action
\be
S=\frac{1}{2\pi\alpha'}\int d\xi^2 \left(\frac{1}{2}\partial_+{\bf x}\partial_-{\bf x}
-\frac{1}{2}f^2 {\bf x}^2 + [{\rm fermions}]\right)
\l{GSpp}\ee
where the fermionic term contains the proper RR-vertex at zero momentum.
The simplicity of the above $\sigma$-model makes the theory then tractable.

The additional importance of studying this background is that it can be 
obtained as a particular contraction of the AdS$\times$S
\cite{maldacena,blau} and therefore its analysis is linked to the 
general issue of holographic gauge/string correspondence.
This point has been addressed in \cite{maldacena}, where the spectrum of 
strings in the pp-wave is recontructed from ${\cal N}=4$ SYM in four dimension
and is a presently studied subject \cite{tutti}.

In this framework, 
there appears a strong resemblance in the string spectrum recontruction 
by the gauge theory operators with
the world-sheet generation in Matrix String Theory (MST).
This suggests a deeper relation within the holographic approach
between the string-bits and the four dimensional gauge theory.

Moreover, the problem of MST in the pp-wave is interesting also by itself, since 
no curved background has been implemented in the MST framework so far.

A possible strategy to obtain a matrix model for strings on the pp-wave
might be to compactify along an isometric direction the eleven dimensional 
matrix models for pp-waves.
This compactification is rather troubling because of the (space-)time dependence of
the supersymmetry parameters and therefore the results obtainable this way
need to be carefully analized from the point of view of supersymmetry.

An alternative strategy for model building, which will be developed in this letter, is a
constructive one. We will first study deformations of ten dimensional super
Yang-Mills (SYM) \cite{sym} induced by the addition of a quadratic term in the spinorial field --
accompanied by a suitable purely bosonic term -- in the action
and then see under dimensional reduction to two dimensions which kind of deformed matrix string
models are available.

As a counter-check we study the deformations of matrix theory induced by
this mechanism via dimensional reduction to one dimension
and we find the pp-wave martix theory of \cite{CLP}
as a particular solution of more general possible structures.

In the next section we study the possible deformations
of {\cal N}=1 SYM in ten dimensions
induced by the addition of a constant bi-spinorial coupling
as $\int \bar\Psi H \Psi$ (plus purely bosonic terms)
with a particular attention to the surviving supersymmetries.
In the next section we perform a dimensional reduction from ten to one
dimension of the results of the previous section and find
the relevant matrix models for particles.
In the third section we face the problem of a formulation of matrix 
string theory on pp-wave background and find the existence of a plethora
of new models with supernumerary supersymmetries.
The last section is left for the conclusions and some open questions.
A deformation of the IKKT matrix model \cite{IKKT} is also discussed in
Appendix A.

\section{Deformation of SYM via a constant bispinor}

In this section we describe a general deformation scheme for ${\cal N}=1$
supersymmetric Yang-Mills theories by applying it to the ten dimensional case.
This deformation is induced by the addition to the action of a purely quadratic
fermionic term plus some bosonic completition.
Here we study the general equations for the deformed supersymmetry.

The action of ${\cal N}=1$ D=10 SYM theory with gauge group $U(N)$ is 
\be
S^0_{10}=\frac{1}{g_{10}^2}\int d^{10}x Tr\left(-\frac{1}{4}F^2+\frac{i}{2}
\Psi^T\Gamma^0\Gamma^\mu D_\mu\Psi\right)
\l{dieci}\ee
where $\Psi$ is a Weyl-Majorana spinor and in the adjoint representation of
the $U(N)$ gauge group.
The action is invariant under the following supersymmetries
\be
\delta^0A_\mu=\frac{i}{2}\bar\epsilon\Gamma_\mu\Psi
\qquad
\delta^0\Psi=-\frac{1}{4}\Gamma^{\mu\nu}F_{\mu\nu}\epsilon
\l{susy0}\ee
\be
\delta^{0'}A_\mu=0 \qquad \delta^{0'}\Psi=\epsilon'
\l{susy0'}\ee
where $\epsilon$ and $\epsilon'$ are {\it constant} Weyl-Majorana spinors.

Notice that, if we perform a variation with $\epsilon$ and $\epsilon'$ 
{\it non constant} Weyl-Majorana spinors, we get
\be
\delta S^0_{10}=\frac{1}{g_{10}^2}\int d^{10}x Tr \bar\Psi
\left\{
-\frac{i}{4} F_{\mu\nu}\Gamma^{\mu\nu}\Gamma^\rho \partial_\rho\epsilon
-i\Gamma_\rho F^{\rho\mu}\partial_\mu\epsilon
+i\Gamma^\mu\partial_\mu\epsilon'
\right\}
\l{invece}\ee
which is linear in the spinor field $\Psi$.

To deform the theory, we add a fermionic bilinear to the ten dimensional action as
$$         
S_{H}=\frac{1}{g_{10}^2}\int d^{10}x \frac{i}{2}Tr \Psi^T\Gamma^0 H \Psi
$$
where $H$ is a real chiral bispinor such that $H=-\Gamma_{11}H=H\Gamma_{11}$
and such that $\Gamma^0H$ is anti-symmetric.
Because of the above consistency conditions, the expansion 
of $H$ in $\Gamma$ matrices correspond just to a three form (plus its dual) 
\footnote{This is actually the standard analysis of type I R-R tensor fields.}.

Obviously, this term breaks supersymmetry (\ref{susy0}) and (\ref{susy0'}) 
and our aim is to construct 
deformations of the field variations $\delta^0$ and $\delta^{0'}$ and of the action such that
no other fermionic term enters the action.
A simple degree counting shows that this is possible by adding 
a further purely bosonic term $S_b$ to the action if the new supersymmetry
variation $\delta$ is such that
\be
\delta A_\mu=\delta^0 A_\mu
\quad{\rm and}\quad
\delta \Psi= \delta^0\Psi + K[A]\epsilon
\label{var}\ee
where $K[A]$ is real chiral bispinor such that
$\Gamma_{11}K[A]=K[A]\Gamma_{11}=K[A]$ 
which we take to be a local functional of the bosonic field
$A_\mu$ only. We drop the condition that $\epsilon$ is constant.
In terms of $\Gamma$-matrices analysis, $K[A]$ is given in terms of even
self dual tensors.
The closure of the the deformed susy algebra implies that $K[A]$ is linear
\footnote{More general forms of $K[A]$ could perhaps be considered in an open
superalgebra framework.}in $A_\mu$.

We consider therefore the total action functional 
\be
S=S^0_{10}+S_H+\frac{1}{g_{10}^2}S_b
\l{action}\ee
and we impose the invariance $\delta S=0$ under the transformations (\ref{var}).
The total variation $\delta S$ turns out to be linear in the fermion 
field $\Psi$ and therefore we are left with a single condition
\be
-\frac{1}{4}F^{\mu\nu}\Gamma_{\mu\nu} \Gamma^\rho \partial_\rho\epsilon
-\Gamma_\rho F^{\rho\mu}\partial_\mu\epsilon
-\frac{1}{4} H F^{\mu\nu}\Gamma_{\mu\nu} \epsilon
+\left(\Gamma^\rho D_\rho +H\right) K[A]\epsilon
-\frac{1}{2} \frac{\delta S_b}{\delta A_\mu}\Gamma_\mu\epsilon
=0
\label{inva}\ee
equivalent to $\delta S=0$.

Notice that $S$ is always invariant under $\delta'=\delta^{0'}$
if $\epsilon'$ satisfies the condition
$\left(\Gamma^\mu\partial_\mu + H\right)\epsilon'=0$.
Further supersymmetries are then classified by the possible solutions of eq.(\ref{inva}).

Notice that if we would require ten dimensional Lorentz invariance, we would
get $H=0$ and therefore this deformation is naturally suitable for dimensionally reduced
theories, where the Lorentz invariance along the reduced dimensions becomes 
R-symmetry which will be eventually broken\footnote{
Anyhow, the kind of breaking of the full ten dimensional Lorentz invariance
induced by the presence of the bi-spinorial $\bar\Psi H\Psi$ term
resembles a higher rank breaking dual to the one induced by non
commutativity parameters.} together with some of the original supersymmetries.
In studing eq. (\ref{inva}), we will assume that $K[A]$ and $S_b$ are 
respectively gauge covariant (a section of the adjoint gauge bundle) 
and gauge invariant with respect to the gauge group of the dimensionally reduced theory.

Equations analogous to (\ref{inva}) can be worked out for reductions of deformed lower
dimensional {\cal N}=1 SYM theories and/or for SYM theories coupled with matter.

\section{Matrix Theory on the pp-wave}

Matrix theory \cite{bfss} can be obtained by dimensional reduction of
$SYM_{10}$ along nine space dimensions.
There has been recently \cite{maldacena,CLP} proposed a model for matrix theory on 
pp-wave backgrounds in eleven dimension.

Here we show how the model there obtained follow as a particualr cases 
from dimensional reducion along nine space directions of the deformed 
ten dimensional SYM that we studied in the previous section. 

In particular we show that all the set of supersumerary supergravity solutions 
studied in \cite{CLP} are contained in a larger class of matrix models.

Let us split the ten dimensional index as $\mu=0,I$, where $I=1,\dots,9$.

We parametrize $K[A]=K^IA_I$ and consider the following bosonic functional
$$
S_b=\frac{1}{g^2}\int dx^0 Tr\left\{
Q^{IJ}A_ID_0A_J+ M^{IJ}A_IA_J+N^{IJK}A_I\left[A_J,A_K\right]
\right\}
$$
where $Q$ and $N$ are completely antisymmetric and $M$ is symmetric.
Substituting $S_b$
in the dimensionally reduced (to $1+0$) equations (\ref{inva}),
we get three bispinorial equations, namely
$$
\frac{1}{4}\left[\Gamma_{IJ},H\right]+\frac{1}{2}\left(\Gamma_IK_J-\Gamma_JK_I\right)
+\frac{1}{4}\Gamma_{IJ}L
-\frac{1}{2}\left(3N^{KIJ}\Gamma_K-Q^{IJ}\Gamma_0\right)\left(\frac{1+\Gamma_{11}}{2}\right)=0
$$ $$
-\frac{1}{2}\Gamma^{0I}(H+L)-\frac{1}{2}H\Gamma^{0I}+\Gamma^0K^I
+Q^{IJ}\Gamma^J\left(\frac{1+\Gamma_{11}}{2}\right)=0
$$ \be
\left[HK^I-\Gamma^0K^I\Gamma^0(H+L)-M^{IJ}\Gamma_J\right]\epsilon=0
\l{yyy}\ee
where we parametrized $(\Gamma^0\partial_0+H+L)\epsilon=0$, with $L$ a generic
chiral bispinor.

The first two equations imply 
$$
K^I=-\frac{1}{2}\left(\frac{1}{12}\Gamma^IN_3+\frac{1}{4}N_3\Gamma^I+\Gamma^0Q^{IK}\gamma^K\right)
\left(\frac{1+\Gamma_{11}}{2}\right)
$$
and 
$$
H=\frac{1}{4}\left(N_3+\Gamma^0Q_2\right)\left(\frac{1+\Gamma_{11}}{2}\right)
\quad {\rm and}\quad
L=-\frac{1}{3}N_3\left(\frac{1+\Gamma_{11}}{2}\right)
$$
where $N_3=N^{IJK}\Gamma^{IJK}$ and $Q_2=Q^{IJ}\Gamma^{IJ}$.

The third equation (\ref{yyy}) reduces to
\be
\left\{
\Gamma^I(N_3)^2+6N_3\Gamma^IN_3+9(N_3)^2\Gamma^I
-12\Gamma^0\left(3N_3Q^{IK}\Gamma^K+Q^{IK}\Gamma^KN_3\right)+144\cdot 2M^{IJ}\Gamma^J
\right\}\epsilon=0
\l{stacippa}\ee
which generalizes the structure equation for supernumerary supersymmetries 
of 11 dimensional pp-wave solutions found in \cite{CLP} (namely eq.(20))
to matrix models with couplings to constant magnetic fields $Q$.

The dimensionally reduced action -- obtained by defining $A_I=igX_I$ and
$\Psi=g\Theta$ -- reads
$$
S_{mmm}=S_{mm}+S_m
$$
where
$$
S_{mm}=
\int dx^0 Tr \left\{
-\frac{1}{2}D_0X^ID_0X^I-\frac{g^2}{4}\left([X^I,X^J]\right)^2
+\frac{i}{2}\Theta^tD_0\Theta -\frac{g}{2}\bar\Theta\Gamma^I[X^I,\Theta]
\right\}
$$
is the usual matrix theory action and 
$$
S_m=\int dx^0 Tr \left\{
\frac{i}{8}\bar\Theta \left(N_3+\Gamma^0Q_2\right)\Theta 
-Q^{IJ}X_ID_0X_J- M^{IJ}X_IX_J-igN^{IJK}X_I\left[X_J,X_K\right]
\right\}
$$
is the deformation part.

The whole action $S_{mmm}$ is invariant under the following field transformations 
$$
\delta A_0=\frac{ig}{2}\bar\epsilon\Gamma_0\Theta
\quad
\delta X^I=\frac{1}{2}\bar\epsilon\Gamma^I\Theta
$$
$$
\delta\Theta=\frac{i}{2}D_0X^I\Gamma^{0I}\epsilon+\frac{g}{4}[X^I,X^J]\Gamma^{IJ}\epsilon+
$$ \be
-\frac{1}{2}\left(\frac{1}{12}\Gamma^IN_3+\frac{1}{4}N_3\Gamma^I+\Gamma^0Q^{IK}\gamma^K\right)
X^I\epsilon
\l{mmmsusy}\ee
if the $\epsilon$ spinor satisfies eq.(\ref{stacippa}) and 
$$
\left(\Gamma^0\partial_0+\frac{1}{4}\left(-\frac{1}{3}N_3+\Gamma^0Q_2\right)\right)\epsilon=0
\quad{\rm and}\quad
\left(\Gamma^0\partial_0+\frac{1}{4}\left(N_3+\Gamma^0Q_2\right)\right)\epsilon'=0
$$

Therefore we obtained a generalization of the matrix models for pp-wave of
\cite{CLP} and have shown that all the classification of pp-wave vacua
can be obtained directly from the matrix theory.

Let us notice that the constant magnetic field $Q_{IJ}$ appearing
in $S_{mmm}$ cannot be in general re-absorbed in the other couplings 
by a change of variables.
One can try to re-absorb it by a coordinate change to a co-rotating frame
as
$$
X^I\quad\to\quad \left(e^{-tQ}\right)^{IJ}X^J
\quad {\rm and} \quad
\Theta \quad\to\quad e^{-\frac{1}{4}tQ_2}\Theta
$$
together with a mass matrix redefinition $M^{IJ}\to M^{IJ}-2Q^{IK}Q^{KJ}$.
It turns out however that in the co-rotating frame one finds
a new 3-tensor and a new mass matrix which are generically
time dependent because of the induced rotation.
The time dependence of the new 3-tensor and mass matrix 
can be avoided only if they are invariant under the rotation
generated by the anti-symmetric matrix $Q$.

Therefore we conclude that the model $S_{mmm}$ effectively
generalizes the one studied in \cite{CLP} if 
$[N_3,Q_2]\not=0$ or $QM+MQ\not=0$ which hold 
in the generic case.

\section{Matrix String Theory on the pp-wave}

Type IIA Matrix String Theory \cite{etc,MST} can be obtained by reducing the ${\cal N}=1$ D=10 SYM
theory with gauge group $U(N)$ down to two dimensions \cite{sym}. 
MST is directly formulated for the IIA string (and the heterotic as well),
while it is just indirectly linked to the type IIB string by T-duality.
We will follow a constructive procedure by looking for possible
supersymmetric deformations of the original
model induced by the presence of a constant additional fermionic bilinear.
This can be done just specifying the general equation obtained above in the
appropriate dimensionally reduced framework.

We split the ten dimensional index $\mu=(\alpha,I)$, with $\alpha=0,9$ and
$I=1,\dots,8$ and solve the dimensionally reduced version of equation (\ref{inva}).

By imposing two dimensional Lorentz invariance and two dimensional gauge
covariance, we find the most general cubic bosonic additional action
functional to be
$$
S_b=\int d^2x Tr \left\{
F_{09}A_Iv_I+\frac{1}{2}M_{IJ}A_IA_J+\frac{1}{3}N^{IJK}A_I[A_J,A_K]
\right\}
$$

We find the following four equations 
$$
\Gamma^\gamma\Gamma^{09}\partial_\gamma\epsilon+H\Gamma^{09}\epsilon+v_1\epsilon=0
$$ $$
-\frac{1}{2}\Gamma^\gamma\Gamma^{\alpha I}\partial_\gamma\epsilon
+\Gamma^\alpha K^I\epsilon-\frac{1}{2}H\Gamma^{\alpha I}\epsilon
+\frac{1}{2}v^I\epsilon^{\alpha\beta}\Gamma_\beta\epsilon=0
$$ $$
-\frac{1}{4}\Gamma^\gamma\Gamma^{IJ}\partial_\gamma\epsilon
+\frac{1}{2}\left(\Gamma^IK^J-\Gamma^Jk^I\right)\epsilon
-\frac{1}{4}H\Gamma^{IJ}\epsilon
-\frac{1}{2}N^{IJK}\Gamma^{K}\epsilon=0
$$ \be
\Gamma^\gamma K^I\partial_\gamma\epsilon+HK^I\epsilon-\frac{1}{2}
M^{IJ}\Gamma^J\epsilon=0 \l{uffa}\ee

The first three equations are equivalent to
$$
K^I=-\frac{1}{2}\left(\frac{1}{12}N_3\Gamma^I+\frac{1}{4}v_1\Gamma^I\Gamma^{09}+v^I\Gamma^{09}\right)
\quad{\rm and}\quad
H=\frac{1}{12}N_3-\frac{1}{4}v_1\Gamma^{09}
$$
and the following differential equation for the spinor $\epsilon$
\be
\partial_\alpha\epsilon+\frac{1}{2}\Gamma_\alpha\left(\frac{1}{12}N_3+\frac{3}{4}v_1\Gamma^{09}\right)\epsilon=0
\l{spin}\ee
while the last equation (\ref{uffa}) is
the mass/flux equation
$$
\left(\frac{1}{12}N_3\Gamma^I-\frac{1}{4}v_1\Gamma^I\Gamma^{09}-v^I\Gamma^{09}\right)\cdot
\left(\frac{1}{12}N_3+\frac{3}{4}v_1\Gamma^{09}\right)
+$$
\be
+\left(\frac{1}{12}N_3\Gamma^I-\frac{1}{4}v_1\Gamma^{09}\right)
\left(\frac{1}{12}N_3+\frac{1}{4}v_1\Gamma^I\Gamma^{09}+v^I\Gamma^{09}\right)
+
M^{IJ}\Gamma^J=0
\l{m/f}\ee

Moreover the integrability condition for (\ref{spin}), which is given by
\be 
\left(N_3+9v_1\Gamma^{09}\right)^T
\left(N_3+9v_1\Gamma^{09}\right)\epsilon=0
\l{inte}\ee
has to be considered.

The reduced action then reads
\be
S_{mms}=S_{ms}+S_m
\l{ppmst}\ee
where 
$$
S_{ms}=
\int d^2x Tr \left\{
-\frac{1}{4g^2}F^{\alpha\beta}F_{\alpha\beta}+
\frac{1}{2}\eta^{\alpha\beta}D_\alpha X^ID_\beta X^I-\frac{g^2}{4}\left([X^I,X^J]\right)^2
+\right. $$ $$\left.
+\frac{i}{2}\bar\Theta\Gamma^\alpha D_\alpha\Theta -\frac{g}{2}\bar\Theta\Gamma^I[X^I,\Theta]
\right\}
$$
is the usual matrix string theory action and 
$$
S_m=\int d^2x Tr \left\{
\frac{i}{8}\bar\Theta \left(\frac{1}{3}N_3-v_1\Gamma^{09}\right)\Theta +
\frac{i}{g}F_{09}X_Iv_I-\frac{1}{2}M_{IJ}X_IX_J-\frac{ig}{3}N^{IJK}X_I[X_J,X_K]
\right\}
$$
is the massive/flux part.

The action (\ref{ppmst}) is invariant under the following field variations
$$
\delta A_\alpha=\frac{ig}{2}\bar\epsilon\Gamma_\alpha\Theta
\quad
\delta X^I=\frac{1}{2}\bar\epsilon\Gamma^I\Theta
$$
$$
\delta\Theta=\left(-\frac{1}{4g}F_{\alpha\beta}\Gamma^{\alpha\beta}+\frac{g}{4}[X^I,X^J]\Gamma^{IJ}
-\frac{i}{2}D_\alpha X^I\Gamma^{\alpha I} +
\right.$$
\be \left.
-\frac{1}{2}\left(\frac{1}{12}N_3\Gamma^I+\frac{1}{4}v_1\Gamma^I\Gamma^{09}+v^I\Gamma^{09}\right)
X^I \right)\epsilon+\epsilon'
\l{ppmstss}\ee
provided eqs.(\ref{spin}), (\ref{m/f}) and (\ref{inte}) hold for $\epsilon$
and $\epsilon'$ satisfies
$$
\left(\Gamma^\alpha\partial_\alpha+\frac{1}{12}N_3-\frac{1}{4}v_1\Gamma^{09}\right)\epsilon'=0
$$

Let us discuss the strong coupling limit as $g\to\infty$ of the above model.
Considering the potential terms $g^2[X,X]^2$ and $g\Theta[X,\Theta]$,
we find that at strong coupling all the fields are projected to a mutually 
commuting subset and the gauge group action becomes 
a parametrization of this choice. Once the choice of a Cartan subalgebra $T$ is
made, the leftover gauge group is generically $U(1)^N$ semidirect with the 
Weyl group $S_N$.

At strong coupling the physical gauge fields degrees of freedom decouple
from the other fields and the curvature field $\frac{F_{09}}{g}=F$ plays the role of an
auxiliary field
\footnote{Notice that in two dimensions a gauge field has only a finite number
of degrees of freedom \cite{hoso} whose role in matrix string theory has been 
studied in \cite{loriano}. We disregard these finite degrees of freedom for
the time being.}  
.
The action at strong coupling is therefore (all the fields are now Cartan valued)
$$
S_{mms}^\infty=
\int d^2x Tr_T \left\{
+\frac{1}{2} F^2+
\frac{1}{2}\eta^{\alpha\beta}\partial_\alpha X^I\partial_\beta X^I
+\frac{i}{2}\bar\Theta\Gamma^\alpha \partial_\alpha\Theta 
\right.+
$$ \be
\left.
\frac{i}{8}\bar\Theta\left(\frac{1}{3}N_3-v_1\Gamma^{09}\right)\Theta +
iF X_Iv_I-\frac{1}{2}M_{IJ}X_IX_J
\right\}
\l{sca}\ee
and the action is invariant under the transformations 
$$
\delta F=\frac{i}{2}\left[\partial_0\left(\bar\epsilon\Gamma_9\Theta\right)
-\partial_9\left(\bar\epsilon\Gamma_0\Theta\right)\right]
\quad
\delta X^I=\frac{1}{2}\bar\epsilon\Gamma^I\Theta
$$
$$
\delta\Theta=-\frac{1}{2}\left(F\Gamma^{09}+
\partial_\alpha X^I\Gamma^{\alpha I} +
\left(\frac{1}{12}N_3\Gamma^I+\frac{1}{4}v_1\Gamma^I\Gamma^{09}+v^I\Gamma^{09}\right)
X^I \right)\epsilon+\epsilon'
$$
under the same constrains as before.

Integrating out the auxiliary field $F$, we get a light-cone string action
for the (matrix) strings given by
$$
S=\frac{1}{2}
\int d^2x Tr_T \left\{
\eta^{\alpha\beta}\partial_\alpha X^I\partial_\beta X^I
+i\bar\Theta\Gamma^\alpha \partial_\alpha\Theta+
\frac{i}{4}\bar\Theta\left(\frac{1}{3}N_3-v_1\Gamma^{09}\right)\Theta+
\right.$$ $$\left.
-\left(M_{IJ}-v_Iv_J\right)X_IX_J
\right\}
$$
This result is obtained as strong coupling limit of the classical action and
it should be checked if there are quantum corrections adding further
complicancies. These are absent in the flat matrix string theory case because
of strong non renormalization theorems for the theory with the full $(8,8)$
supersymmetry. It would be important to verify if these non renormalization
properties are still effective for the models we have been considering here.
The role of the abovementioned discrete gauge theory degrees of freedom should also be
considered in detail.

\subsection{A set of models}

A particular set of Matrix String Theories in pp-wave backgrounds
can be given for example as follows.

We solve the constraints (\ref{m/f}) and (\ref{inte})
with the choice
$$
N_3=9\mu\eta\Gamma^{123}
\quad{\rm and}\quad
v_1=\mu\Gamma^4
$$
where $\mu$ is a mass parameter and $\eta=\pm 1$.

Moreover we break $1/4$ of the original dymamical supersymmetry, by choosing
$$
(\Gamma^{09}+\eta')\epsilon=0
\quad{\rm and}\quad
(\Gamma^{1234}-\eta\eta')\epsilon=0
$$
where $\eta'=\pm 1$.
The two flags $\eta$ and $\eta'$ label with their values four different
models.

The mass matrix $M^{IJ}$ turns out diagonal
\be
M=\frac{\mu^2}{16}\left[\left(3\eta'(3+\eta)-4\right){\bf 1}_3
\oplus 12(1+\eta'){\bf 1}_1 \oplus 2(3\eta'+1){\bf 1}_4\right]
\l{masspart}\ee
and these models all break the original $SO(8)$ R-symmetry of the flat ($\mu=0$)
model to $SO(3)\times SO(4)$.
The full action is
$$ S^{\eta\eta'}_{mms}= S_{ms}+ S^{\eta\eta'}_{m} $$
where $S_{ms}$ is the flat Matrix String action of the previous section
and $S^{\eta\eta'}_{m}$ is given by
$$
S^{\eta\eta'}_{m}=
\int d^2x Tr \left\{
\frac{i\mu}{2}\bar\Theta \left[\frac{3}{4}\eta\Gamma^{123}-\frac{1}{4}\Gamma^{409}\right]
\Theta +\frac{i\mu}{g}F_{09}X^4
-\frac{ig\mu\eta}{2}\sum_{I,J,K=1}^3\varepsilon^{IJK}X_I[X_J,X_K]
\right. $$ \be \left.
-\frac{\mu^2}{32}\left[(3\eta'(3+\eta)-4)\sum_{I=1}^3(X^I)^2+12(1+\eta')(X^4)^2
+2(3\eta'+1)\sum_{I=5}^8(X^I)^2\right]
\right\}
\l{massactpart}\ee

The deformed supersymmetry transformations are then
$$
\delta A_\alpha=\frac{ig}{2}\bar\epsilon\Gamma_\alpha\Theta
\quad
\delta X^I=\frac{1}{2}\bar\epsilon\Gamma^I\Theta
$$
$$
\delta\Theta=\left(-\frac{1}{4g}F_{\alpha\beta}\Gamma^{\alpha\beta}+\frac{g}{4}[X^I,X^J]\Gamma^{IJ}
-\frac{i}{2}D_\alpha X^I\Gamma^{\alpha I} +
\right.$$ \be \left.
-\frac{1}{2}\left(\frac{3}{4}\mu\eta
  \Gamma^{123}\Gamma^I+\frac{1}{4}\mu\Gamma^4\Gamma^I\Gamma^{09}
+\mu\delta^{I4}\Gamma^{09}\right)
X^I \right)\epsilon+\epsilon'
\l{ppmstsspart}\ee
and the action is left invariant provided the above conditions on $\epsilon$
are satisfied and $\epsilon'$ satisfies
$$\left[\Gamma^\alpha\partial_\alpha
+\frac{\mu}{4}\left(3\eta\Gamma^{123}-\Gamma^{409}\right)\right]\epsilon'=0
$$

\section{Conclusions and open questions}

In this letter we provided a set of possible models for Matrix String Theory
on different pp-wave backgrounds and a general scheme to analyse their
supersymmetries.
It will be of interest to see how they will enter the gauge/string
correspondence (a role played by matrix string theory has been recently
advocated in \cite{gopa}). 

A crucial check of MST is the interpretation of the inverse gauge coupling 
as the actual type IIA string coupling.
This was obtained in \cite{MST} by a carefull analysis of the supersymmetry
preserving vacua. These are classified my solutions of an integrable system,
the Hitchin equations, whose spectral surfaces materialize the interacting 
worldsheet diagrams in the strong coupling expansion.
A proper index counting on these spectral surfaces
provides in the relevant path integrals
the crucial factor of $(1/g)^{-\chi}$, where
$\chi$ is the Euler characteristic of the string worldsheet,
and this implies the interpretation of $\frac{1}{g}$ as the string coupling $g_s$.
Therefore, it is crucial to study susy-preserving configurations
to check if and how the presence of the pp-wave background
changes the reconstruction of the superstring interaction.
This will hopefully provide a clear scheme --
which is {\it intrinsic} the Matrix String Theory hypothesis --
for the study of closed 
string interaction in a pp-wave background.

Let us notice that we can perform a full dimensional 
reduction to zero dimensions while keeping our scheme 
(at the price of excluding at least some kinematical supersymmetries) 
and get a deformed version of the IKKT matrix model \cite{IKKT}.
This is done in Appendix A.
It would be nice to understand how to interpret this deformation  
from the type IIB string point of view of \cite{IKKT}.

As it is evident from the results obtained here, the number of constrains on
the surviving supersymmetries increases with the number of unreduced
dimensions.
It would be interesting to study also deformations of higher dimensional SYM
theories along the paths of the general scheme presented here
to see if some of them still can have some supernumerary supersymmetries.

{\bf Acknowlodgements}.
I would like to thank M.Bertolini and L.Bonora for discussions.
Work supported by the European Community's Human Potential
Programme under contract HPRN-CT-2000-00131 Quantum Spacetime
in which G.B. is associated to Leuven.

\appendix
\section{Deforming the IKKT model}

The deformation of the IKKT model \cite{IKKT} along the scheme outlined in this letter
shows a peculiar phenomenon of a complete breaking of the kinematical 
supersymmetries $\epsilon'$.
In this appendix we present the formulas for the relevant deformation.

We perform dimensional reduction of the ten dimensional gauge theory down to
zero dimensions and we choose the bosonic completition of the action 
to be 
$$
S_b= Tr\left\{ M^{\mu\nu}A_\mu A_\nu +N^{\mu\nu\rho}A_\mu[A_\nu,A_\rho]\right\}
$$
and find the following two equations (we have $K[A]=K^\mu A_\mu$)
$$
\left\{
H\Gamma^{\mu\nu}-2\left(\Gamma^\mu K^\nu-\Gamma^\nu K^\mu\right)
+6N^{\mu\nu\rho}\Gamma_\rho
\right\}\epsilon=0
$$
\be
\left\{HK^\mu-M^{\mu\nu}\Gamma_\nu\right\}\epsilon=0
\l{app1}\ee

The first one is identically solved by 
$$
K^\mu=-\frac{1}{16}\Gamma^\mu N_3-\frac{1}{8}N_3\Gamma^\mu
\quad {\rm and}\quad
H=\frac{1}{4}N_3
$$
where, as usual, $N_3=N_{\mu\nu\rho}\Gamma^{\mu\nu\rho}$, while
and the last one out of (\ref{app1}) is left as a mass/flux equation
\be
\left[
N_3\left(\Gamma^\mu N_3+2N_3\Gamma^\mu\right)+4^3M^{\mu\nu}\Gamma_\nu
\right]\epsilon=0
\l{app2}\ee

The deformed IKKT matrix model is given by the action matrix function
$$
S^{IKKT}_m= S^{IKKT}+S_m
$$
where
$$
S^{IKKT}=Tr\left\{-\frac{g^2}{4}\left[X^\mu,X^\nu\right]^2-\frac{g}{2}
\bar\Theta\Gamma_\mu\left[X^\mu,\Theta\right]\right\}
$$
is the IKKT matrix model for the flat IIB Schild action and 
$$
S_m=Tr\left\{\frac{i}{8}\bar\Theta N_3\Theta-
M^{\mu\nu}X_\mu X_\nu +igN^{\mu\nu\rho}X_\mu[X_\nu,X_\rho]\right\}
$$
is the deformation part.

The total action is invariant under the transformations
\be
\delta X^\mu=\frac{1}{2}\bar\epsilon\Gamma^\mu\Theta
\quad {\rm and}\quad
\delta \Theta= \frac{g}{4}\left[X^\mu,X^\nu\right]\Gamma_{\mu\nu}\epsilon
-\frac{i}{16}X^\mu\left(\Gamma_\mu N_3+2N_3\Gamma_\mu\right)\epsilon
\l{app3}\ee
if (\ref{app2}) is satisfied.
Notice the fact that all the kinematical linear supersymmetries $\epsilon'$ have been
at least partly lost due to the constrain $N_3\epsilon'=0$.

The easiest realization of these models can be obtained by 
studying the case
$N_3=\mu\Gamma^{789}$
for which, choosing the mass matrix to be
$$
M=-\frac{\mu^2}{4^3}\left({\bf \eta}_7\oplus 3{\bf 1}_3\right)
$$
we solve identically in $\epsilon$ the mass/flux constrain (\ref{app2})
and we retain the whole 16 $\epsilon$ supersymmetries -- all the
kinematical ones being lost --
while breaking the ten dimensional Lorentz group $SO(1,9)$ to $SO(1,6)\times SO(3)$.

\end{document}